\documentclass[prl,twocolumn,superscriptaddress,showpacs]{revtex4}
\usepackage{amsmath}
\usepackage{bm}
\usepackage{graphicx}
\usepackage{amssymb}
\usepackage{datetime}
\begin{document}
\newcommand{\figdir}{.}
\newcommand{\figwidth}{0.9\columnwidth}
\newcommand{\ffigwidth}{0.4\columnwidth}
\newcommand{\warwick}{Department of Physics and Centre for Scientific Computing, University of Warwick, Coventry, CV4 7AL, United Kingdom}
\newcommand{\osaka}{Department of Physics, Graduate School of Science, Osaka University, 1-1 Machikaneyama, Toyonaka, Osaka 560-0043, Japan}
\title{Critical parameters from generalised multifractal analysis at the Anderson transition}
\author{Alberto Rodriguez}
\email[Corresponding author:\\]{A.Rodriguez-Gonzalez@warwick.ac.uk}
\affiliation{
Department of Physics and Centre for Scientific Computing, University of Warwick, Coventry, CV4 7AL, United Kingdom}
\author{Louella J. Vasquez}
\affiliation{
Department of Physics and Centre for Scientific Computing, University of Warwick, Coventry, CV4 7AL, United Kingdom}
\author{Keith Slevin}
\affiliation{
Department of Physics, Graduate School of Science, Osaka University, 1-1 Machikaneyama, Toyonaka, Osaka 560-0043, Japan}
\author{Rudolf A. R\"omer}
\affiliation{
Department of Physics and Centre for Scientific Computing, University of Warwick, Coventry, CV4 7AL, United Kingdom}
\date{$Revision: 1.50 $, compiled \today, \currenttime}
%
\begin{abstract}
We propose a generalization of multifractal analysis that is applicable to the critical {\em regime} of the Anderson localization-delocalization transition. The approach reveals that the behavior of the probability distribution of wavefunction amplitudes is sufficient to characterize the transition.
In combination with finite-size scaling, this formalism  permits the critical parameters to be estimated without the need for conductance or other transport measurements.
Applying this method to high-precision data for wavefunction statistics obtained by exact diagonalization of the three-dimensional Anderson model, we estimate the critical exponent $\nu=1.58\pm 0.03$.
\end{abstract}
%
\maketitle


The statistical analysis of spatial probability and density fluctuations, which has a distinguished history \cite{Man82}, has recently received new impetus.
Scanning-tunnelling spectroscopy now allows the direct measurement of the spatial variation of charge densities \cite{MorKMG02}.
Dramatic advances in cold atom physics are stimulating the study of Anderson localization in Bose-Einstein condensates by imaging of the atomic densities. This permits the observation of the exponential decay of the wavefunctions and direct measurement of localization lengths \cite{BilJZB08}. Anderson-type transitions can now be investigated experimentally in quasi-periodic disorder potentials \cite {RoaDFF08} and in cold-atom realizations of the kicked-rotor \cite{LemCSG09}. Similarly, the spatial localization of light \cite{WieBLR97} has recently been studied in nano devices with slow-wave structures \cite{MooPYB08}.
The fundamental tool to characterize density fluctuations and the scale invariance of spatial distributions at critical points is multifractal analysis (MFA).
Recently, a predicted symmetry  of the multifractal spectrum at the Anderson transition \cite{MirFME06} has been confirmed by experimental studies of vibrations in elastic networks \cite{FaeSPL09}.
MFA has also furnished insights into the theoretical foundations of the quantum Hall transition \cite{EveMM08a}.
However, the application of MFA is restricted to the critical point, where the relevant probability distributions are truly multifractal \cite{Jan94a,Jan98}.
Up to now, this has meant that additional computer simulations or experiments must be performed in advance to locate the critical point precisely, since any error here will adversely affect the results.

\begin{figure}
  \centering
   \includegraphics[width=0.9\columnwidth]{\figdir/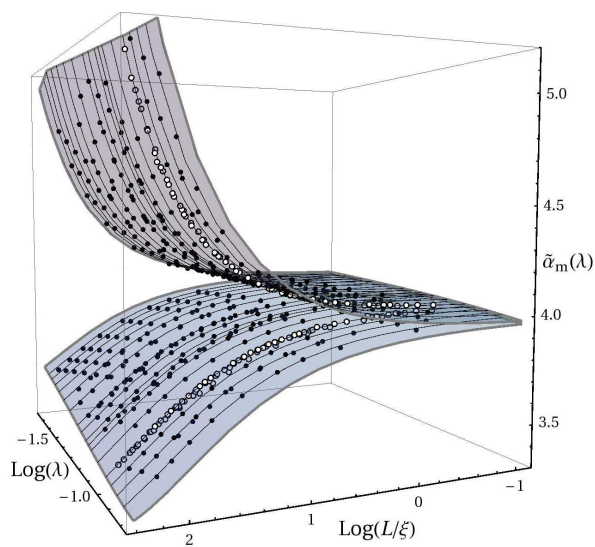}
   \caption{(color online) {\em Two-parameter} finite-size scaling result for $\widetilde{\alpha}_{\rm m}(\lambda)$ at energy $E=0$ when fitted according to Eq.~\eqref{eq-alphaq} indicated by the shaded surface. Data ($\bullet$, $\circ$) and fit parameters are as described in the second row and caption of Table \ref{tab-3Dfits}. The black lines on the surface highlight the different values of $\lambda$ [symbol ($\circ$) highlights $\lambda=0.1$].
   The estimated $\alpha_0=4.09$ corresponds to the extrapolation of $\widetilde{\alpha}_{\rm m}(\lambda)$ as $\lambda\rightarrow 0$.}
\label{fig-pq-fss-3d}
\end{figure}

In this Letter we propose a generalized MFA that is applicable throughout the {\em critical regime} and not just at the critical point.
Our approach is motivated by the behavior of the probability density function (PDF) of the wavefunction intensities, $\mathcal{P}(\widetilde{\alpha};W,L,\lambda)$. Here, $\widetilde{\alpha} \equiv \ln\mu_k/\ln \lambda$ with $\mu_k\equiv \sum_{i=1}^{\ell^d} |\psi_i|^2$ the summed wavefunction probability in the $k$-th cubic box of linear size $\ell$ in a lattice with volume $L^d$, and $\lambda\equiv\ell/L$.
%
%
\begin{figure}
  \centering
  \includegraphics[width=.98\columnwidth]{\figdir/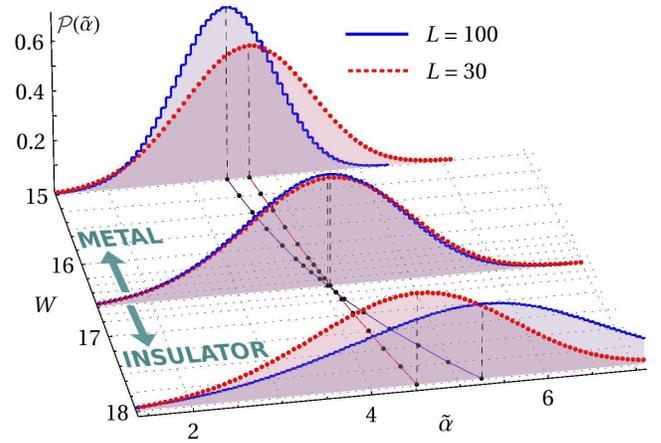}
   \caption{(color online) Evolution of the wavefunction amplitude distributions $\mathcal{P}(\widetilde{\alpha};W,L)$ as a function of disorder $W$ across the Anderson transition, at {\em fixed} $\lambda=0.1$ for two system sizes $L$.
   Each distribution has been computed with $10^4$ wavefunctions.
   The data points ($\bullet$) and solid lines on the bottom plane mark the trajectories of the maximum $\widetilde{\alpha}_{\rm m}$.
   For clarity, distributions are shown at $W=15$, $16.6$ and $18.0$ only.}
\label{fig-pdf-3d}
\end{figure}
%
We have applied our method to the three-dimensional Anderson model with box-distributed site energies of width $W$. We have calculated more than $1.5$ million uncorrelated wavefunctions by exact diagonalization of system sizes up to $100^3$ \cite{VasRR08a}. We find that the parameter dependence of $\mathcal{P}(\widetilde{\alpha};W,L,\lambda)$, as displayed in Fig.~\ref{fig-pdf-3d}, is sufficient to characterize the Anderson transition. For fixed $\lambda$, the distribution becomes scale invariant at the critical point and away from the transition its maximum, $\widetilde{\alpha}_{\rm m}$, exhibits finite size scaling (FSS) behavior: $\widetilde{\alpha}_{\rm m}$ shifts in opposite directions in the different phases at a rate which depends on $L$. This provides an alternative way to estimate the critical parameters of the transition that is not based on transport properties such as the conductance.
We believe that our approach is particularly valuable in experiments where the PDF of wavefunction amplitudes is accessible, e.g., through LDOS measurements using STM techniques  \cite{MorKMG02} or in ultracold Bose/Fermi gases in disordered optical lattices \cite{BilJZB08,RoaDFF08}.
For the Anderson model our estimates of $W_{\rm c}$ and $\nu$ in Table \ref{tab-tq-alphaq-fits} are in excellent agreement with previous transfer matrix results \cite{SleO99a}, which resolves the long-standing issue of systematically smaller exponents found in previous diagonalization studies \cite{ZhaK97}.
\begin{table}[tb]
\begin{tabular}{ccllccccc}
\hline\hline
        & $\lambda$ & $\nu$ & $W_{\rm c}$ & $N_D$ & $N_P$ & $\chi^2$ & $p$ & $n_0 n_1 m_{\varrho}m_{\eta}$  \\
\hline
 $\widetilde{\tau}_2$ & 0.1 & 1.58(52,66) & 16.57(50,61) & 153 & 13 & 151 & 0.2 & 5 1 3 0 \\
 $\widetilde{\tau}_2$ & 0.2 & 1.59(57,61) & 16.56(52,58) & 153 & 12 & 158 & 0.2  & 6 0 2 0\\
 $\widetilde{\tau}_3$ & 0.1 & 1.62(57,66) & 16.56(42,61) & 153 & 10 & 133 & 0.7 & 5 0 1 0 \\
 $\widetilde{\alpha}_{\rm m}$ & 0.1 & 1.56(54,59) & 16.53(49,55) & 153 & 10 & 131 & 0.7  & 3 2 1 0 \\
 $\widetilde{\alpha}_1$ & 0.1 & 1.61(58,64) & 16.56(52,59) & 81 & 7 & 89 & 0.1  & 2 0 1 0 \\
\hline
\hline
\end{tabular}
\caption{The estimates of the critical parameters $\nu$ and $W_{\rm c}$, together with 95\% confidence intervals, from {\em one-parameter} FSS. The number of data is $N_D$, the number of parameters is $N_P$, $\chi^2$ is the value of the chi-squared statistic for the best fit, and $p$ is the goodness of fit probability. The orders of the expansions are specified in the last column. The system sizes are $L\in[20, 100]$, and the range of disorder is $W\in\left[15,18\right]$ except for $\widetilde{\alpha}_1$ where $W\in\left[16,17\right]$. The data uncertainty is on average 0.6\% for $\widetilde{\tau}_3$, 0.3\% for $\widetilde{\tau}_2$ and 0.1\% for $\widetilde{\alpha}_1$ and $\widetilde{\alpha}_{\rm m}$.%
}
\label{tab-tq-alphaq-fits}
\end{table}
In addition, the size dependence of the PDF on $L$ {\em and} $\lambda$ (or, equivalently, $\ell$) suggests the possibility of two-parameter FSS. This is illustrated in Fig.~\ref{fig-pq-fss-3d} where scaling is shown as a function of both $L/\xi$ and $\lambda$, with $\xi$ the correlation/localisation length.


Our generalization of MFA starts by considering the $q$-moments of the wavefunctions defined as $R_q\equiv\sum_k \mu_k^q$, where the sum runs over the $(L/\ell)^d$ boxes of linear size $\ell$.
At criticality and in the thermodynamic limit ($L\rightarrow\infty$), due to the multifractal nature of the states, the only relevant parameter is $\lambda\equiv \ell/L$ and the moments scale as $\langle R_q \rangle \underset{\lambda\rightarrow 0}{\sim} \lambda^{\tau_q}$. Here, the brackets denote an average over disorder. Away from the transition, however, the moments depend on $\ell$, $L$ and the disorder $W$.
It follows from scaling arguments that close to the transition
$
 R_q = R_q (L/\xi,\lambda) \equiv R_q(L/\xi,\ell/\xi)
$ \cite{YakO98},
with $\xi\equiv\xi(W)$ diverging at the critical point $W_{\rm c}$ as $\xi \propto |W-W_{\rm c}|^{-\nu}$.
Close to criticality we can write
$
 \langle R_q \rangle (W,L,\lambda)= \lambda^{\tau_q} \mathcal{R}_q\left( L/\xi,\lambda\right),
$
which can be rearranged as follows,
\begin{equation}
   \widetilde{\tau}_q(W,L,\lambda) = \tau_q + \frac{q(q-1)}{\ln \lambda} {\mathcal{T}_q}\left( L/\xi,\lambda\right).
   \label{eq-tauq}
\end{equation}
Here, ${\mathcal{T}_q}$ is related to the original $\mathcal{R}_q$ and we have defined a generalized mass exponent as
$\widetilde{\tau}_q(W,L,\lambda)\equiv \ln \langle R_q \rangle/\ln \lambda$ which becomes the usual $\tau_q$ at $W_{\rm c}$ and in the limit $\lambda\rightarrow 0$. The factor $q(q-1)$ has been explicitly included to satisfy $\widetilde{\tau}_0=\tau_0\equiv -d$ and $\widetilde{\tau}_1=\tau_1\equiv 0$. From Eq.\ \eqref{eq-tauq} it is straightforward to obtain the scaling law for the singularity strengths $\widetilde{\alpha}_q\equiv d \widetilde{\tau}_q/dq$,
\begin{equation}
 \widetilde{\alpha}_q(W,L,\lambda) = \alpha_q + \frac{1}{\ln \lambda} \mathcal{A}_q \left( L/\xi,\lambda\right),
   \label{eq-alphaq}
\end{equation}
where  the second term on the rhs will be non-zero for all $q$ values, and the generalized exponents are defined as
$\widetilde{\alpha}_q(W,L,\lambda)\equiv \langle \sum_k \mu_k^q \ln \mu_k \rangle/ \left(\langle R_q \rangle \ln \lambda\right)$.
Consequently, we can define a $W$, $L$ and $\lambda$ dependent generalized singularity spectrum $\widetilde{f}_q\equiv q\widetilde{\alpha}_q - \widetilde{\tau}_q$, obeying
\begin{equation}
 \widetilde{f}_q(W,L,\lambda) = f_q + \frac{q}{\ln \lambda} \mathcal{F}_q \left( L/\xi,\lambda\right).
 \label{eq-falphaq}
\end{equation}
Eqs.~\eqref{eq-tauq}--\eqref{eq-falphaq} suggest a wide range of generalized exponents that can be used to perform FSS and obtain $W_{\rm c}$ and $\nu$. In addition, the scale invariant multifractal exponents $\tau_q$, $\alpha_q$ and $f_q$ at the critical point can be estimated from the same FSS study {\em without} the need to know $W_{\rm c}$ beforehand. Moreover, the use of different moments of the wavefunctions provides a test of the stability of the estimates for the critical parameters, as these should be $q$-independent.

The generalized multifractal spectrum \eqref{eq-falphaq} is related to the PDF of the wavefunction amplitudes as $\mathcal{P}(\widetilde{\alpha};W,L,\lambda)\propto \lambda^{d - \widetilde{f}(\widetilde{\alpha};W,L,\lambda)}$.
 As we approach the thermodynamic limit ($\lambda\rightarrow 0$) at the critical point, this becomes the usual relation $\mathcal{P}_\lambda(\alpha)\propto \lambda^{d - f(\alpha)}$ \cite{RodVR09}.
As shown in Fig.\ \ref{fig-pdf-3d}, for fixed $\lambda$, the PDF becomes scale invariant at the transition.
The generalized exponents can also be calculated from the distribution $\mathcal{P}(\widetilde{\alpha};W,L,\lambda)$, which may be useful when the wavefunctions cannot be probed individually and only partial information about the PDF is accessible.
For $q=0$, we have $\widetilde{\alpha}_0\equiv \langle \lambda^d \sum_k \ln \mu_k/\ln\lambda\rangle = \langle \widetilde{\alpha} \rangle$, which corresponds to the mean value of the PDF. When $L\rightarrow\infty$,  $\langle \widetilde{\alpha} \rangle$ converges towards the position of the maximum of the PDF $\widetilde{\alpha}_{\rm m}$
\footnote{The PDF is not symmetric around its maximum \cite{RodVR09} and hence $\widetilde{\alpha}_{m} \neq \langle \widetilde{\alpha}\rangle$ ($ \equiv \widetilde{\alpha}_0$) in general for finite $\lambda$. But both quantities agree in the limit $L\rightarrow\infty$.}.
While $\widetilde{\alpha}_{\rm m}$ and $\widetilde{\alpha}_{0}\equiv \langle \widetilde{\alpha} \rangle$ may differ quantitatively at finite $L$, they obey the same scaling law with the same critical parameters. Therefore the scaling of either the mean value or the position of the maximum of the PDF as a function of $W$, $L$ and $\ell$ may be used to estimate the critical parameters.
We note that the scaling law \eqref{eq-alphaq} that we give here for our generalized multifractal exponents $\widetilde{\alpha}_0$ and $\widetilde{\alpha}_{\rm m}$ is different from the scaling laws suggested in the past \cite{Jan94a,HucS92}.

We first present results for standard one-parameter ($L/\xi$) FSS at {\em fixed} $\lambda$ values \cite{SleO99a,MilRSU00}.
Let $\Gamma_{\lambda}(W,L)$ denote either $\widetilde{\tau}_q(W,L,\lambda)$ or $\widetilde{\alpha}_q(W,L,\lambda)$. We introduce a set of fit functions which include two kinds of corrections to scaling, (i) nonlinearities of the $W$ dependence of the scaling variables and (ii) an irrelevant scaling variable that accounts for a shift of the disorder value at which the $\Gamma_{\lambda}(W,L)$ curves cross. We use
$
  \Gamma_{\lambda}(W,L)={\mathcal G}(\varrho L^{1/\nu}, \eta L^{-|y|})
$,
where ${\mathcal G}$ denotes the rhs of either Eq.\ \eqref{eq-tauq} or \eqref{eq-alphaq} and $\varrho$ and $\eta$ are the relevant and irrelevant scaling fields, respectively. The function $\Gamma_{\lambda}(W,L)$ is expanded to first order in the irrelevant scaling variable as
$\Gamma_{\lambda}(W,L)= {\mathcal G}_0(\varrho L^{1/\nu}) +\eta L^{-|y|} {\mathcal G}_1(\varrho L^{1/\nu})$,
and subsequently
${\mathcal G}_s= \sum_{k=0}^{n_s} a_{sk} \varrho^k L^{k/\nu}$.
The fields $\varrho$ and $\eta$ are expanded in terms of $w\equiv(W_{\rm c}-W)/W_{\rm c}$ up to order $m_{\varrho}$ and $m_{\eta}$, respectively, such that
$
  \varrho(w)=\sum_{m=1}^{m_{\varrho}} b_m w^m$,
   $\eta(w)=\sum_{m=0}^{m_{\eta}} c_m w^m
$,
with $b_1=c_0=1$. The expansions of the fit functions are truncated at orders
$n_0, n_1, m_{\varrho}, m_{\eta}$.
The orders of these expansions should be kept as low as possible,
while giving an acceptable goodness of fit probablity $p$.
We emphasize that the results of the FSS analysis are valid only if the goodness-of-fit is acceptable.

\begin{figure}
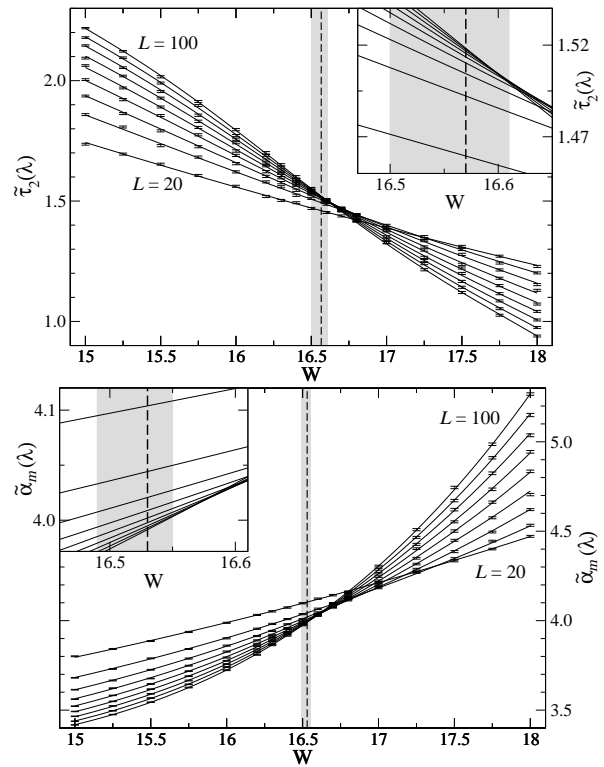

  \centering
  \includegraphics[width=.9\columnwidth]{\figdir/fig-tau2-lambda0.1.eps}
  \includegraphics[width=.9\columnwidth]{\figdir/fig-alpha0-lambda0.1.eps}
   \caption{Plot of $\widetilde{\tau}_2$ (top) and $\widetilde{\alpha}_{\rm m}$ (bottom) at $E=0$ and $\lambda=0.1$ as a function of disorder at various system sizes $L\in[20,100]$. The errorbars denote standard deviations obtained from averaging over disorder. The lines are plotted  according to Eq.~\eqref{eq-tauq} and \eqref{eq-alphaq}, respectively, for the fit parameters as in the first and fourth rows of Table \ref{tab-tq-alphaq-fits} and with irrelevant exponents $y=-1.8\pm 0.4$ and $y=-1.65\pm 0.12$. The vertical dashed lines show the estimated $W_{\rm c}$ and its confidence interval is indicated by the grey region. The insets show the FSS functions in an enlarged view of the critical region.}
\label{fig-tau-alpha-fss}
\end{figure}

In Fig.~\ref{fig-tau-alpha-fss} (top)
we show a fit for $\widetilde{\tau}_2$ data at $\lambda=0.1$.
As $L$ increases, the generalised exponent $\widetilde{\tau}_2$ approaches the metallic and insulating limits correponding to $d(q-1)$ for $W<W_{\rm c}$ and $0$ for $W>W_{\rm c}$, respectively.
We emphasize that the value of $\widetilde{\tau}_2$ at the critical point is $\lambda$-dependent, and it is only in the limit $\lambda\rightarrow 0$ that $\widetilde{\tau}_2(\lambda)$ converges towards the scale invariant $\tau_2$.
A similar behavior is observed in Fig.~\ref{fig-tau-alpha-fss} (bottom) for $\widetilde{\alpha}_{\rm m}$, the position of the maximum of the PDF.
In this case
$\lim_{L\rightarrow\infty} \widetilde{\alpha}_{\rm m} = d $ in the metallic side and $\lim_{L\rightarrow\infty} \widetilde{\alpha}_{\rm m} = \infty $ in the insulating phase.
The trajectories of $\widetilde{\alpha}_{\rm m}$ as a function of disorder for different $L$ are also shown on the bottom plane of Fig.\ \ref{fig-pdf-3d}.
The PDF $\mathcal{P}(\widetilde{\alpha};W,L)$ was obtained from the numerical histogram of wavefunction intensities \cite{RodVR09}.
The position of its maximum was estimated by fitting $\mathcal{P}(\widetilde{\alpha})$ to $\lambda^{h(\widetilde{\alpha})}$ where $h(\widetilde{\alpha})$ is a polynomial, which allows for the non-symmetric and non-Gaussian nature of the distribution \cite{RodVR09}.
The precision of the $\widetilde{\alpha}_{\rm m}$ data was determined by performing the fit $100$ times on independent distributions obtained from subsets of $100$ states each for every set $(W,L)$.

In Table \ref{tab-tq-alphaq-fits}, we show representative results for $W_{\rm c}$ and $\nu$ from $3$ fits of $\widetilde{\tau}_q$ and $2$ fits for $\widetilde{\alpha}_q$ at various $\lambda$ and $q$ values.
The analysis is based on a total of $1,530,000$ wavefunctions generated at energy $E=0$ for system sizes $L^3$ in the range $20^3$ to $100^3$, and disorder values $W$ from $15$ to $18$, where for each pair $(W,L)$ we average over $10^4$ independent states.
Table \ref{tab-tq-alphaq-fits} shows that all fits give estimates of $\nu$ which (i) are consistent with each other, (ii) agree with the transfer-matrix-method results $\nu=1.57(55,59)$ \cite{SleO99a} and $\nu=1.62(55,69)$ \cite{MilRSU00} and (iii) are significantly larger than $1$ and, within the accuracy, different from $1.5$ \cite{Gar08}. Our results also agree with the estimation of $\nu$ obtained from the quantum kicked rotor which was recently realized experimentally using cold atoms \cite{LemCSG09}.
The large irrelevant shift of $W_{\rm c}$ seen in Fig.~\ref{fig-tau-alpha-fss}
is comparable to those observed for higher Lyapunov exponents \cite{SleO01}. This might explain the variation in the estimated value of $\nu$ from previous works based on exact diagonalization \cite{ZhaK97}, as only the use of very large system sizes can resolve this shift unambiguously.

The two-parameter scaling suggested in Eqs.\ \eqref{eq-tauq} and \eqref{eq-alphaq} and shown in Fig.\ \ref{fig-pq-fss-3d} is based on a scaling function of the type
\begin{equation}
\mathcal{A}_q(L/\xi,\ell/\xi)= \mathcal{A}_q^0(\varrho L^\frac{1}{\nu},\varrho \ell^\frac{1}{\nu}) +\eta \ell^{-|y|} \mathcal{A}_q^1(\varrho L^\frac{1}{\nu} ,\varrho \ell^\frac{1}{\nu}),
\label{eq-2par-sf}
\end{equation}
for $\widetilde{\alpha}_q$ (similarly for $\widetilde{\tau}_q$) where the dominant irrelevant scaling is determined by $\ell$
\footnote{At fixed $\lambda$ the irrelevant component turns into $L^{-|y|}$ as used before.}.
The functions $\mathcal{A}_q^s$ are expanded in their arguments and relevant/irrelevant fields, and the expansion is characterized by the indices $n_L^0$, $n_\ell^0$, $n_L^1$, $n_\ell^1$, $m_{\varrho}$ and $m_{\eta}$.
The two-parameter scaling provides a simultaneous estimation of the critical parameters $W_{\rm c}$, $\nu$ and the scale invariant multifractal exponents $\tau_q$, $\alpha_q$.
As shown in Table \ref{tab-3Dfits}, the estimated values for $W_{\rm c}$ and $\nu$ from two-parameter FSS, using a large number of data for integer and non-integer $q$, are in agreement with those obtained at fixed $\lambda$ (Table \ref{tab-tq-alphaq-fits}).
\begin{table}[tb]
\begin{tabular}{lllccccc}
\hline\hline
        & $\nu$ & $W_{\rm c}$ & $N_D$ & $N_P$ & $\chi^2$ & $p$ & expansion  \\
\hline
 $\widetilde{\tau}_2$ & 1.56(52,60) & 16.57(55,59) & 544 & 20 & 558 & 0.15 & 3\,2\,0\,1\,3\,0 \\
 $\widetilde{\alpha}_{\rm m}$ & 1.56(55,58) & 16.55(54,56) & 544 & 16 & 494 & 0.85 & 2\,2\,0\,2\,1\,0 \\
 $\widetilde{\tau}_{1.1}$ & 1.60(55,64) & 16.57(55,59) & 224 & 11 & 232 & 0.17 & 2\,1\,0\,0\,1\,0 \\
 $\widetilde{\alpha}_1$ & 1.60(55,64) & 16.56(54,58) & 224 & 13 & 230 & 0.18 & 3\,1\,0\,0\,1\,0 \\
\hline
\hline
\end{tabular}\caption{The estimates of $W_{\rm c}$ and $\nu$ from {\em two-parameter} FSS. Labels are described in Table \ref{tab-tq-alphaq-fits}.
The last colum shows the orders of the expansion: $n_L^0$, $n_\ell^0$, $n_L^1$, $n_\ell^1$, $m_{\varrho}$ and $m_{\eta}$. The scale invariant multifractal exponents obtained from the fits are $\tau_2=1.21(20,22)$, $\alpha_0=4.09(08,10)$, $\tau_{1.1}=0.184(183,185)$ and $\alpha_1=1.93(92,94)$ respectively. The data used correspond to  $L\in[20, 100]$, $\ell\geqslant 2$ satisfying $0.02\leqslant\lambda\leqslant 1/7$ and $W\in[15,18]$, except for the last two where $W\in[16.2,16.8]$.}
\label{tab-3Dfits}
\end{table}
The data involved in this analysis may differ in $\ell$ but share the same $L$, hence there is a certain degree of correlation that may affect the fit \cite{WeiJ09}. However, the values of $\nu$ in Table \ref{tab-3Dfits} are within the accuracy the same as those of the uncorrelated FSS in Table \ref{tab-tq-alphaq-fits} \footnote{Using the covariance matrix of the data in $\chi^2$ minimization we find that the value of $\nu$ is unaffected by the correlations.}.
In Fig.~\ref{fig-pq-fss-3d} we show the scaling for $\widetilde{\alpha}_{\rm m}$. The shaded surface denotes Eq.\ \eqref{eq-alphaq} using the two-parameter scaling function \eqref{eq-2par-sf} with the irrelevant correction subtracted,
displayed as a function of $L/\xi$ and $\lambda$, where the correlation length is given by $\xi=|\varrho|^{-\nu}$.
The scaling function exhibits an upper and a lower sheet populated by values of $\widetilde{\alpha}_{\rm m}$ corresponding to extended ($W<W_{\rm c}$) and localised ($W>W_{\rm c}$) states respectively. The merging of the two sheets as $\xi\rightarrow\infty$ at constant $\lambda$ determines the estimation of $\nu$. The additional extrapolation of the merging point as $\lambda\rightarrow 0$ gives the scale invariant $\alpha_0$.


In conclusion, we have proposed a generalisation of multifractal concepts such as mass exponents,
singularity strengths and the multifractal spectrum that is applicable to the critical {\em regime} of the Anderson transition.
The combination of the generalized MFA with FSS provides the critical parameters of the transition and enables MFA to be applied
without knowing the exact position of the critical point in advance.
We have tested our method on the Anderson model of an electron in a disordered system, and
we estimate the critical exponent that describes the divergence of the localization length to be
$\nu=1.58 \pm 0.03$, in agreement with previous transfer matrix calculations \cite{SleO99a}.
The method is applicable to other models with critical fluctuations and to the wealth of such experimental data that is now becoming available \cite{MorKMG02,BilJZB08,RoaDFF08,MooPYB08}.

\begin{acknowledgements}
 The authors gratefully acknowledge EPSRC (EP/F32323/1, EP/C007042/1, EP/D065135/1) for financial support. A.R.\ acknowledges financial support from the Spanish government (FIS2009-07880). R.A.R.\ thanks T.\ Ohtsuki for an inspiring discussion about this topic in 2002.
\end{acknowledgements}

\end{document}